\documentclass[prd,a4paper,superscriptaddress,nofootinbib,10pt]{revtex4} 
\usepackage{graphicx}
\usepackage{amssymb}
\usepackage{amsmath}
\usepackage{lscape}
\usepackage{mathrsfs}
\usepackage{textcomp}
\usepackage{epsfig}
\usepackage{slashed}
\usepackage{comment}
\renewcommand{\d}{\mathrm{d}}

\begin{document}
\title{Quantum space and quantum completeness}

\author{Tajron Juri\'c}
\email{tjuric@irb.hr}
\affiliation{Rudjer Bo\v{s}kovi\'c Institute, Bijeni\v cka  c.54, HR-10002 Zagreb, Croatia}
\affiliation{Instituto de Fisica, Universidade de Brasilia,
Caixa Postal 04455, 70919-970, Brasilia, DF, Brazil}

\date{\today}

\begin{abstract}

Motivated by the question whether quantum gravity can  ``smear out'' the classical singularity we analyze a certain quantum space and its quantum-mechanical completeness. 
Classical singularity is understood as a geodesic incompleteness, while quantum completeness requires a unique unitary time evolution for test fields propagating on an underlying background. 
Here the crucial point is that quantum completeness renders the Hamiltonian (or spatial part of the wave operator) to be essentially self-adjoint in order to generate a unique time evolution.
We examine a model of quantum space which consists of a noncommutative BTZ black hole probed by a test scalar field. We show that the quantum gravity (noncommutative) effect is to enlarge the 
domain of BTZ parameters for which the relevant wave operator is essentially self-adjoint. This means that the corresponding quantum space is quantum complete for a larger range of BTZ parameters rendering the conclusion that 
in the quantum space one observes the effect of ``smearing out'' the singularity. 

\end{abstract}

\maketitle

\section{Introduction}

In general relativity (GR) it is often to encounter solutions with singularity, i.e. black holes. The presence of singularity is equivalent to the statement that the corresponding metric is geodesically incomplete.
Geodesics are generalization of straight lines in curved space, and therefore describe the motion of the free falling test particles.  A more precise statement is the following:  a spacetime is called geodesically complete if every maximal geodesic is defined on an entire real line. If a spacetime is (time-like) geodesically incomplete, then the evolution of certain test particles is not well defined after a finite proper time and is said to be singular \cite{hawking}. \\

There is a great hope that classical singularities could be ``smeared out'' when considering quantum theory (and quantum gravity) \cite{Ashtekar:2003hd, Ashtekar:2005qt, Bojowald:2006rp, Yonika:2017qgo, Saini:2016vgo, Saini:2014qpa, Greenwood:2008ht, Wang:2009ay}. There are examples of static spacetimes with time-like singularities where a quantum test particle is completely well behaved for all time. \\ 

How can one formulate a condition in quantum theory  which would determine whether or not a certain spacetime is singular? One way is to look at the expectation value of certain ``physical operators'' and see whether they diverge or not. Another way is to analyze under which condition is the evolution of any state uniquely defined for all time. Such system is non-singular or quantum-mechanically complete. If this is not the case, then one looses predictability and the system is said to be singular. \\

A simple example of a classically incomplete, but quantum-mechanically complete space is given by a non-relativistic particle on a bounded interval. This system is classically singular because the associated ``spacetime'' is geodesically incomplete. One can define the Hamiltonian $H$ to be the Laplacian acting on a wave functions that vanish smoothly at the boundary. This operator is symmetric, but not yet self-adjoint. There are many so-called self-adjoint extensions (SAE) of this operator, where changing the boundary condition one enlarges the domain of the operator rendering it self-adjoint. One of these extensions must be chosen in order to evolve quantum states. This is analogous to a typical time-like singularity in GR, where one also has to chose the appropriate boundary condition at the singularity. In both of these cases the evolution is not unique until some extra information is specified.\\

Another example is the classical motion on a half-line. This motion is classically complete at the end point if there are no initial conditions such that  the trajectory runs off to the end point in a finite time. The classical motion is complete at the end point if the potential grows unbounded from above near the end point \cite{reedsimon}. More precisely,  in quantum mechanics on a half line, a time-independent potential is called quantum mechanically complete, if the associated Hamiltonian is essentially self-adjoint on the space of $\mathcal{C}^{\infty}_{0}(\left< 0,+\infty \right>)$-functions of compact support on the half-line with the origin excluded \cite{reedsimon}.\\

Hydrogen atom is a very good example of a singular classical theory which is nonsingular quantum-mechanically. Since the Coulomb potential is bounded from above near the origin, the electron can reach the origin in a finite time, therefore the classical motion of electron in the Coulomb potential is classically incomplete. However, when Coulomb potential is probed by the quantum electron (via its bound state) it becomes quantum complete. In other words, the classical singularity of the Coulomb potential is not reflected in any observable related to the bound-state electron.\\

What is the physical reason for the difference between classical and quantum completeness? The reason for this difference is that these spacetimes produce an effective repulsive barrier which shields their classical singularity, so that the   quantum wave packets bounce off this barrier. From this viewpoint, geodesics correspond to the geometric optics limit of finite waves. Only in this unphysical limit one can reach the singularity. There is a direct analog of this in singular geometries. \\

Let us consider quantum mechanics of a free particle moving on an $n+1$ dimensional Riemannian manifold $(\mathcal{M}, g)$. Here the Hilbert space consists of square integrable functions on $\mathcal{M}$, with a measure given by the proper volume element $d^{n+1}x\sqrt{-g}$. The Hamiltonian is given by the Laplacian on the manifold. It is known that for Riemannian manifolds $(\mathcal{M}, g)$ which are geodesically complete, the corresponding Laplacian is essentially self-adjoint(it has a unique SAE \cite{chernof}). This means that when a space is classically nonsingular, then it is also quantum mechanically complete. \\

What happens when the metric is geodesically incomplete? Can we still have a unique self-adjoint Laplacian? The answer is positive \cite{horowitz}. One can illustrate this by examining a spherical symmetric metric
\begin{equation}\label{met}
ds^2=dr^2+f^2(r)d\Omega_n
\end{equation}
where $d\Omega_n$ is the standard metric on the $n$-sphere. The domain of Laplacian naturally consists of smooth functions with compact support away from the origin. In order to see the self-adjoint property of the Laplacian it is sufficient to consider the eigenvalue equation for the Laplacian with purely imaginary eigenvalues\footnote{see $Theorem \  X.2$ in \cite{reedsimon}}, i.e.
\begin{equation}
\Delta_n \psi=\pm i\psi
\end{equation}
and show that there are no square integrable solutions \cite{reedsimon}. Loosely speaking, the above equation will expose to us all the domains in which the Laplacian is for sure not self-adjoint, since one of the most important properties of self-adjoint operators is that they have real eigenvalues. Now, using the separation of variables $\psi=R(r)Y(\Omega)$ one obtains the radial equation
\begin{equation}
R^{\prime\prime}+\frac{nf^{\prime}}{f}R^{\prime}-\frac{c}{f^2}R=\pm i R
\end{equation}
where $c\geq 0$ is an eigenvalue of the Laplacian on the $n$-sphere and prime is the derivative with respect to the radial coordinate $r$. The self-adjointness is  equivalent to the statement that for each $c$ and each choice of the eigenvalue $\pm i$ , there are no square-integrable solutions near the origin. To see this, one can look only at the case $c=0$, since $c>0$ only increases the divergence of the solution at the origin\footnote{Note that the term $\pm i R(r)$ is negligible near the origin.} $r=0$. If $f(r)=r^k$ near the origin, and the solution is given by $R(r)=r^{\alpha}$ (where $\alpha=1-nk$) we see that there are no square-integrable solutions (with respect to the measure $d^{n+1}x\sqrt{-g}\longrightarrow r^{kn}dr$) if 
\begin{equation}
k\geq\frac{3}{n}.
\end{equation}
therefore, we come to conclusion that any metric of the form \eqref{met} and $f(r)=r^k$ ($k\geq\frac{3}{n}$) near the origin is quantum-mechanically nonsingular. We know that the metric \eqref{met} is geodesically incomplete unless $k=1$, so we see that even in this simple example we found a large class of classically singular geometries, which are quantum-mechanically complete. 

In \cite{horowitz} one can find various examples of  geodesically incomplete static space-times, with time-like curvature singularities, which are quantum mechanically complete. Their work stimulated a lot of research concerning classically incomplete but quantum-mechanically complete spacetime \cite{isibasi, isibasi1, konkowski}.\\

It is known that in static globally hyperbolic spacetimes one can 
construct a consistent quantum field theory (QFT) \cite{aschtecar}. This is due to 
the fact that in such spaces it is possible to define a consistent 
single-particle quantum theory, and then each one-particle state is 
equal to the corresponding classical field, which enables the 
construction of QFT. Furthermore, following the works of Wald \cite{wald} , 
the authors of \cite{horowitz} were able to show that the aforementioned is 
possible even in certain spacetimes with time-like singularities. The 
key point was to prove that the problem of defining the time-evolution 
of a Klein-Gordon (KG) field in an arbitrary static spacetime (with 
singularities) can be reformulated as a problem of constructing SAE of 
the spatial part of the wave operator. In \cite{helfer, hofmann} it was discussed 
that for a general time-dependent spacetime the only adequate 
description is in terms of QFT. This required the study of evolution 
of classical fields in a singular background so that the dynamical 
spacetime was treated as an external background. In such a setting 
unitarity is no longer required and the notion of quantum completeness 
needs to be appropriately reconsidered \cite{hofmann}.\\

Another approach in dealing with the singularities is presented in \cite{Kraus}. In this paper the BTZ black hole singularities and horizons are described in terms of AdS/CFT amplitudes, where the naive divergences associated with the Milne type singularity is regulated by an $i\epsilon$ prescription. Singularities can also be avoided in black hole solutions coming from String theory using the so called fuzzball proposal \cite{Mathur}.\\

In this paper we will examine a model of quantum space and its quantum completeness. For the model we will use a noncommutative (NC) BTZ black hole \cite{ncbtz, ncbtz1, ncbtz2, ncbtz3}. The noncommutativity is inspired by the idea that general relativity and Heisenberg’s uncertainty principle together imply that the spacetime has a noncommutative structure \cite{dop1, dop2}. In such NC spaces one abandons the idea of smooth manifolds and replaces the 
algebra of smooth functions with some NC algebra for which it is possible to develop the whole differential calculus \cite{connes}. We will show that the NC structure of spacetime will improve the quantum completeness with regards to the commutative case. Namely, after one examines the KG equation in this NC setting we see that the main NC effect is to widen the range of BTZ parameters for which the spatial part of the Laplacian is essentially self-adjoint, rendering unique time evolution.\\

The paper is organized as follows. In section II we present the BTZ black hole and investigate the dynamics of a test scalar field. After using von-Neumann's method we find the criterion on BTZ parameters that lead to a quantum complete spacetime. In section III we first sketch the NC setting from our previous papers \cite{ncbtz, ncbtz1, ncbtz2, ncbtz3}, and then proceed with the same analysis as in section II. We see that the range of BTZ parameters rendering the quantum completeness is enlarged. Finally, we conclude with some final remarks and future perspectives in section IV.\\

\section{BTZ black hole and the propagation of a scalar  test field}

Probing black holes (BH) with scalar fields reveals important information about thermodynamical properties of the system, such as black hole entropy and Hawking radiation \cite{gthooft, bombelli, thooft1, kumar1, kumar2}. In this approach BH geometry is an external background on which scalar field is analyzed. There are potential problems with such analysis:
\begin{itemize}
\item behavior of the scalar field at the event horizon may not be well defined
\item free energy may diverge due to infinite numbers of modes contributing to it near horizon \cite{gthooft}
\item certain spacetimes are not globally hyperbolic and that leads to difficulties in predicting the field propagation \cite{wald}
\end{itemize}
One possible solution is to impose suitable boundary conditions on the scalar field. For example, in the so called ``brick-wall'' method one imposes that the scalar field vanishes near (on a finite distance from) horizon \cite{gthooft, thooft1}. This way one obtains a well defined free energy and entropy. The choice of boundary condition is not unique and physical results often depend on this particular choice. It is relevant to classify all possible boundary conditions that could lead to a unique well-defined scalar field propagation and to investigate how this choice effects the physical quantities of interest. The study of all possible choices of boundary conditions and their consequences for the scalar field propagation on BTZ black hole was reported in \cite{kumarnormal}. There they used the method of deficiency index due to von Neumann \cite{reedsimon}. They found that for certain range of the system parameters the evolution in not unique, i.e. the space is quantum-mechanically incomplete.  In that case there exists a one-parameter family of SAE of the corresponding KG operator. \\

\subsection{Klein-Gordon equation on BTZ background}
BTZ is a solution of (2+1)-dimensional Einstein equation with negative cosmological constant $\Lambda=-\frac{1}{l^2}$. The BTZ black hole can be obtained by the discrete quotienting of the universal covering space of $AdS_3$, which has a time-like spatial infinity. This feature leads to the lack of global hyperbolicity for BTZ. In \cite{wald, satoh} this issue was handled by  requiring that the solution of KG equation vanishes sufficiently rapidly at the spatial infinity.

The massive spinless BTZ black hole is described by the metric\footnote{Here we use the coordinate system $(t,r,\phi)$, signature $(+,-,-)$ and the natural system of units $G=\hbar=c=1$. } \cite{banados}
\begin{equation}\label{btzmetric}
g_{\mu\nu}=\begin{pmatrix}
\frac{r^2}{l^2}-M&0&0\\
0&-\frac{1}{\frac{r^2}{l^2}-M}&0\\
0&0&-r^2\\
\end{pmatrix},
\end{equation}
where we have taken  the angular momentum to be zero, i.e.  
$J=0$ and  $l$ is related to the cosmological constant $\Lambda$ as $l = \sqrt{-\frac{1}{\Lambda}}$. The KG equation for a massless scalar field in the BTZ background is given by
\begin{equation}
\Box \phi=\frac{1}{\sqrt{\left|g\right|}}\partial_{\mu}\left(\sqrt{\left|g\right|}g^{\mu\nu}\partial_{\nu}\phi\right)=0
\end{equation}
After using separation of variables $\phi=e^{-iEt}e^{im\varphi}R(r)$ we obtain the following 
radial equation\footnote{Here we have used the unique selfadjoint conditions for the angle part $\phi(\varphi)=\phi(\varphi+2\pi)$. It is important to stress that these boundary conditions emerge by demanding that the solution is a single-valued function. It is easy to check, using the von Neumann criterion \cite{reedsimon}, that the angular part of the operator in question is essentially self-adjoint with a unique SAE. This is the reason why in the later parts of this work we will be always concentrated just on the radial part of the equations.} 
\begin{equation}\label{radial}
r\left(M-\frac{r^2}{l^2}\right)\frac{\partial^2 R}{\partial r^2}+\left(M-\frac{3r^2}{l^2}\right)
\frac{\partial R}{\partial r}+\left(\frac{m^2}{r}-E^2\frac{r}{\frac{r^2}{l^2}-M}\right)R=0,
\end{equation}
where $m\in\mathbb{Z}$ is the azimuthal quantum  number.
Now, the eq. \eqref{radial} has an exact solution \cite{satoh, kenmoku, kuwata}.
In order to see this, we use the following substitution
\begin{equation}
z=1-\frac{Ml^2}{r^2},
\end{equation}
and re-express eq.\eqref{radial}  as
\begin{equation}
\label{eom}
z(1-z)\frac{\d^2 R}{\d z^2}+ (1-z)\frac{\d R}{\d z} + \left(\frac{A}{z}+B\right)R=0,
\end{equation}
where the constants $A$ and $B$ are
\begin{equation} \label{coefs}
A=\frac{E^2 l^2}{4M}, \quad B=-\frac{m^2}{4M}.
\end{equation}
We see that the eq.\eqref{eom} has regular singular points at $z=0$, $1$ and  $\infty$, which means that around this points we can find a solutions via Frobenious method. 
 The general solution can be written as 
\begin{equation}
R(z)=z^{\alpha}F(z)
\end{equation}
where we recognize that $F(z)$ is the hypergeometric function satisfying
\begin{equation}
z(1-z)\frac{\d^2 F}{\d z^2}+\left[c-(1+a+b)z\right]\frac{\d F}{\d z}-abF=0.
\end{equation}
where
\begin{equation}
a=\alpha+i\sqrt{-B}, \quad b=\alpha-i\sqrt{-B}, \quad c=2\alpha+1
\end{equation}
and 
\begin{equation}
\alpha=i\sqrt{A}
\end{equation}

Various solutions for different boundary conditions have been found in the literature, i.e. normal modes \cite{kenmoku, kuwata, kumarnormal}, quasi-normal modes \cite{Cardoso:2001hn, Birmingham:2001hc}, self-adjoint extension issues \cite{kumarnormal}, etc.

\subsection{Quantum completeness of the BTZ spacetime}
In order to deal with the difficulties arising  from the lack of global hyperbolicity \cite{wald}, one demands that the solution of \eqref{radial} vanishes at spatial infinity. This solution can be then analytically continued to the horizon $r=l\sqrt{M}$ (or $z=0$), where it diverges. In order to regulate this divergence, one usually introduces a ``brick-wall'' parameter $\epsilon$,  
and one requires that the function vanishes at $z=\epsilon$. This boundary condition leads to a well posed problem, and the corresponding thermodynamical quantities can then be evaluated \cite{satoh} (but they depend on $\epsilon$).\\

Now, in order to see if the BTZ spacetime is quantum complete, one has to investigate the self-adjointness of the Hamiltonian in eq.\eqref{radial} or \eqref{eom}. First let us rewrite \eqref{eom} in the Sturm-Liouville form
\begin{equation}\label{sturm}
\frac{\d}{\d z}\left(p(z)\frac{\d R}{\d z}\right)+q(z)R=-\lambda w(z)R
\end{equation}
so that we identify
\begin{equation}
p(z)=z(1-z), \quad q(z)=B, \quad w(z)=\frac{1}{z}, \quad \lambda=A
\end{equation}
We see that $p(z)$, $q(z)$ and $w(z)$ are all continuous on the interval $z\in\left[\epsilon,\  1\right]$  (i.e. $r\in\left\langle l\sqrt{M}, \ \infty\right\rangle$) and $p(z)$ has a continuous derivative. $w(z)$ is usually called  the weight or density function. Now the map
\begin{equation}
\mathcal{L}[f]=-\frac{1}{w(x)}\left(\frac{\d}{\d x}\left[p(x)\frac{\d f}{\d x}\right]+q(x)f\right)
\end{equation}
can be viewed as a linear operator mapping a function $f$ to another function $\mathcal{L}[f]$. This operator when studied in the context of functional analysis is examined through the eigenvalue equation\footnote{Which is equivalent to \eqref{sturm}}
\begin{equation}
\mathcal{L}[f]=\lambda f
\end{equation}
and the Hilbert space $L^2\left([a,b],\ w(x)dx\right)$ with the scalar product
\begin{equation}
\left\langle f,\ g\right\rangle=\int^{b}_{a}f^{*}(x)g(x)w(x)\d x 
\end{equation}
In our case $w(x)=\frac{1}{x}$ and $[a,b]=[\epsilon, 1]$. The operator $\mathcal{L}$ is defined on sufficiently smooth functions which satisfy the boundary condition
\begin{equation}\label{boundary}
f(\epsilon)=0=f(1)  
\end{equation}
and are square-integrable on $L^2\left([a,b],\ w(x)dx\right)$. This way $\mathcal{L}$ gives rise to a symmetric operator satisfying
\begin{equation}
\left\langle f,\ \mathcal{L}[g]\right\rangle=\left\langle \mathcal{L}[f],\ g \right\rangle
\end{equation}
where we used the partial integration twice and the boundary condition\footnote{Actually this is true for a larger class of boundary conditions:
\begin{equation*}\begin{split}
&\alpha_1 f(a)+\alpha_2 f^{\prime}(a)=0, \quad \alpha^{2}_{1}+\alpha^{2}_{2}>0\\
&\beta_1 f(b)+\beta_2 f^{\prime}(b)=0, \quad \beta^{2}_{1}+\beta^{2}_{2}>0
\end{split}\end{equation*}} 
\eqref{boundary}. In order for $\mathcal{L}$ to be a self-adjoint operator one also requires that the domain of $\mathcal{L}$, that is $D(\mathcal{L})=\left\{f\in L^2\left([a,b],\ w(x)dx\right)| f(\epsilon)=f(1)=0 \ \wedge\  f,f^\prime \ \text{are continuous}\right\}$ is equal to the domain of its adjoint $\mathcal{L}^{\dagger}$, i.e. $D(\mathcal{L})=D(\mathcal{L}^{\dagger})$.\\

In order to determine whether the symmetric operator $\mathcal{L}$ is essentially self-adjoint, or if it admits SAE it is enough to investigate the square integrable solutions of the eigenvalue equations with purely imaginary eigenvalues. In doing so we will replace the energy $E$ with imaginary unit $i$, i.e.  $\lambda(E)=A=\frac{E^2 l^2}{4M}\longrightarrow\lambda(\pm i)=\frac{(\pm i)^2 l^2}{4M}=-\frac{l^2}{4M}\equiv\lambda_{\pm}$ and consider the following eigenequation
\begin{equation}\label{imaginarna}
\mathcal{L}_{\pm}[f_{\pm}]=\lambda_{\pm}f_{\pm}
\end{equation}
This way we are searching for a possible domain in which $\mathcal{L}$ could have imaginary eigenvalues, that is we are searching for domains on which $\mathcal{L}$ is definitely not self-adjoint, because a self-adjoint operator has only real eigenvalues. In the theory of SAE developed by von Neumannn \cite{reedsimon} one is interested in the number of linearly independent square integrable solutions with eigenvalues $+i$ and $-i$. These numbers, called deficiency indices $(n_{+}, n_{-})$ determine the following:
\begin{enumerate}
\item if $(n_{+}, n_{-})=(0,0)$, then $\mathcal{L}$ is essentially self-adjoint
\item if $n_{+}=n_{-}\equiv n$, then $\mathcal{L}$ is not self-adjoint, but admits SAE, and the new domain on which it is self-adjoint is given by
\begin{equation}
D^{SAE}(\mathcal{L})=D(\mathcal{L})\oplus\left\{f_{+}+U_{n}f_{-}\right\}
\end{equation}
where $f_{\pm}$ are solutions of \eqref{imaginarna} and $U_n$ is $n\times n$ unitary matrix. 
\item if $n_{+}\neq n_{-}$, then $\mathcal{L}$ does not admits SAE
\end{enumerate}

Now, consider \eqref{imaginarna} for $+i$ case
\begin{equation}
\mathcal{L}_{+}[f_+]=\lambda_{+}f_+ \quad \Longleftrightarrow \quad z(1-z)\frac{\d^2 f_+}{\d z^2}+ (1-z)\frac{\d f_+}{\d z} + \frac{\lambda_+ f_+}{z}+Bf_+=0
\end{equation}
The solution can be written as
\begin{equation}
f_+ (z)=z^\alpha F(z)
\end{equation}
where $F(z)$ satisfies the hypergeometric differential equation
\begin{equation}\label{hyper}
z(1-z)\frac{\d^2 F}{\d z^2}+\left[c-(1+a+b)z\right]\frac{\d F}{\d z}-abF=0.
\end{equation}
where 
\begin{equation}
a=\alpha+i\sqrt{-B}, \quad b=\alpha-i\sqrt{-B}, \quad c=2\alpha+1
\end{equation}
and 
\begin{equation}
\alpha=i\sqrt{\lambda_+}=-\frac{l}{2\sqrt{M}}, \quad B=-\frac{m^2}{4M}
\end{equation}
Since $c\notin\mathbb{Z}$ then there are two linearly independent solutions of \eqref{hyper} on the horizon $z=\epsilon$, given by $F(a,b,c;z)$ and $z^{1-c}F(a+1-c,\ b+1-c,\ 2-c; z)$. So, the full solution is 
\begin{equation}\label{fplus}
f_+ (z)=C_1 z^\alpha F(a,b,c;z)+C_2 z^{-\alpha}F(a+1-c,\ b+1-c,\ 2-c; z)
\end{equation}
Now we are interested in the behavior for\footnote{Notice that in the spatial infinity $z=1$ the solution is regular and it is sufficient to establish the condition for the existance of non-square integrable solutions just on the horizon $z=0$.} $z\longrightarrow 0$. Since we have $F(...,z=0)=1$, \eqref{fplus} reduces to
\begin{equation}\label{limes}
\lim_{z\rightarrow 0} f_{+}(z)\propto C_1 z^\alpha+C_2 z^{-\alpha}
\end{equation}
We are now interested in the situation where there are NO square integrable solutions\footnote{Notice that the radial part of the measure $\sqrt{-g}d^3 x$ on the horizon is proportional to $ rdr\approx dz$. Also notice that since we are analyzing the operator in the equation \eqref{radial}, that is \eqref{eom} with the imaginary eigenvalue, the complete measure is given by $\sqrt{-g}d^3 x$. If one analyzes the operator $\mathcal{L}$ then one uses the measure $w(z)=\frac{1}{z}$, but the relation between the two operators and measures in questions is quite obvious (multiplying/dividing with $w(z)$). }. So we are looking for a condition on $\alpha$ such that $\lim_{z\rightarrow 0} \left|f_{+}(z)\right|^2$ is divergent more then $z^{-1}$. Since $\alpha=-\frac{l}{2\sqrt{M}}<0$, the second term in \eqref{limes} is regular and the divergence can come only from the first term, giving
\begin{equation}
\lim_{z\rightarrow 0} \left|f_{+}(z)\right|^2 \propto z^{2\alpha}
\end{equation}
Therefore, there will be NO square integrable solutions if
\begin{equation}
2\alpha\leq -1
\end{equation}
which then gives a condition on the BTZ parameters
\begin{equation}\label{condition}
\sqrt{M}\leq l
\end{equation}
Analogously, one can prove that for the same condition \eqref{condition} eigenvalue equation $\mathcal{L}_{-}[f_{-}]=\lambda_{-}f_{-}$ has no square integrable solutions. Therefore we obtained for the deficiency indicies  $(n_{+}, n_{-})=(0,0)$ whenever \eqref{condition} holds, rendering the operator $\mathcal{L}$ essentially self-adjoint, so that the propagation of scalar field is uniquely  defined for all times and we conclude that the BTZ spacetime is quantum complete for the parameter range \eqref{condition}.\\

For the BTZ black hole with mass $M$ and intrinsic angular momentum $J$, the study of self-adjoitness is analogous to the derivation presented above and was also analyzed in \cite{kumarnormal}, where the generalized version of the condition \eqref{condition} can be found. In case the condition \eqref{condition} does not hold, there is a one-parameter family of SAE discussed in \cite{kumarnormal}.\\

\section{NC BTZ}
Properties of spacetime at the Planck scale could be very different from what we observe nowadays and what is predicted by GR. Namely, various models including string theory \cite{string}, loop quantum gravity \cite{loop}, and  NC geometry \cite{connes}, suggests that the spacetime might have a certain discrete structure at the quantum gravity scale. Combining together GR and quantum uncertainty principle one can predict a very general class of NC spaces \cite{dop1, dop2, ahluwalia}. Interestingly, one particular Lie algebraic type of NC space, called $\kappa$-Minkowski space \cite{kappa1, kappa2, kappa3} is associated to a variety of black holes at the Planck scale \cite{btzkappa, ohl}.\\

\subsection{Implementing noncommutativity through NC duality}
In a previous set of works \cite{ncbtz, ncbtz1, ncbtz2, ncbtz3}, the properties of NC ($\kappa$-Minkowski) scalar and fermion fields in the background of a BTZ black hole\cite{banados} were investigated. It is established that probing a spinless BTZ black hole with a $\kappa$-Minkowski scalar field is equivalent to probing a spinning BTZ black hole with a commutative scalar field \cite{ncbtz1}. The effective spin of this dual BTZ black hole was obtained from the corresponding black hole entropy \cite{ncbtz1, ncbtz3}, and it was shown that it depends on the NC parameter. Moreover, it captures the backreaction of the NC scalar filed on the BTZ spacetime.\\

In order to illustrate this NC duality we briefly review the results that were obtained previously. Namely, for the massless NC scalar field in the BTZ black hole background with mass $M$ and $J=0$ is described by the wave equation that can be  symbolically written as
\begin{equation} \label{nckg}
  ({\Box_{g'}} + {\mathcal{O}}(a)) \Phi = 0.
\end{equation}
where
\begin{equation}\label{btzm}
g'_{\mu\nu}=\begin{pmatrix}
M - \frac{r^2}{l^2} &0&0\\
0&\frac{1}{\frac{r^2}{l^2}-M}&0\\
0&0&r^2\\
\end{pmatrix},
\end{equation}
and $a=\frac{1}{\kappa}$ is the NC scale\footnote{$\kappa$-Minkowski algebra is defined by \begin{equation*}
\lbrack \hat{x}^{i},\hat{x}^{j}]=0,\quad \lbrack \hat{x}^{0},\hat{x}^{i}]=
\frac{i}{\kappa }\hat{x}^{i},  
\end{equation*} where the parameter $\kappa$ is related to the Planck mass $M_{Planck}$ or some effective quantum gravity scale.}, $\Box_{g'} $ is the Klein-Gordon (KG) operator in the metric \eqref{btzm}, and the second term ${\mathcal{O}}(a)$ is a generic expression representing a whole set of corrections induced by the NC nature of spacetime. In \cite{ncbtz1} it was shown that \eqref{nckg} can be rewritten in the form 
\begin{equation} \label{gkg}
  {\Box_{g}}  \Phi = 0,
\end{equation}
where $\Box_{g}$ is the
KG operator for the metric\footnote{Now there exist an outer and inner radius, $r_+$ and $r_{-}$ defined by $$M^d =\frac{r^{2}_{+}+r^{2}_{-}}{l^2}, \quad J^{d}=\frac{2r_+ r_{-}}{l}.$$ }
\begin{equation}\label{eqbtzmetric}
g_{\mu\nu}=\begin{pmatrix}
 M^d - \frac{r^2}{l^2} - \frac{{(J^{d})}^2}{4 r^2} &0& \frac{-J^d}{2}\\
 0&\frac{1}{\frac{r^2}{l^2} + \frac{{(J^{d})}^2}{4 r^2} -M^d}&0\\
\frac{-J^d}{2} &0& r^2\\
\end{pmatrix},
\end{equation}
The different black hole parameters $M^d$ and $J^d$ in this dual picture indicate that the black hole in the dual setting acquired an angular momentum, i.e. $J=0\rightarrow J^d\neq 0$. Noncommutativity enables the scalar particle to influence the geometry in which it propagates, thus making the instance for the backreaction mechanism in this particular situation.\\

To determine the specific expression for $J^d$, we consider the entropy of the spinless BTZ black hole probed by the NC scalar field \cite{ncbtz}
\begin{equation} 
S^{NC}=\frac{A_0}{4G}\left(1-\Lambda_{NC}\sqrt{M}\frac{8\pi\zeta(2)}{3l\zeta(3)}\right),
\end{equation}
where $A_0=2\pi l\sqrt{M}$ is the area of the spinless BTZ with mass $M$, and $\Lambda_{NC}=-a\beta$, where $\beta$ is related to the choice of the vacuum (realization/ordering\footnote{For more details we refer the reader to  \cite{Meljanac:2007xb, Govindarajan:2008qa, Juric:2013foa}.}). Now since the NC KG equation \eqref{nckg} is identical to the KG equation in the dual picture \eqref{gkg}, we can postulate the equivalence of entropy
\begin{equation}\label{S}
S^{NC}=S^{d}
\end{equation}
where
\begin{equation}
S^d=\frac{A^d}{4G}, \quad A^d=2\pi r_+ , 
\end{equation}
where $S^d$ is the entropy of the dual BTZ black hole and 
\begin{equation} \label{rplus} 
 r_{+}=\frac{l\sqrt{M^d}}{\sqrt{2}}\sqrt{1+\sqrt{1-\frac{(J^d)^2}{M^2l^2}}},
\end{equation}
is the outer horizon of the dual BTZ black hole. Upon choosing $M^d =M$ equation \eqref{S} gives  the following expression for the induced black hole spin
\begin{equation}\label{J}
(J^d(a))^2=\Lambda_{NC}\frac{64}{3}\pi\frac{\zeta(2)}{\zeta(3)}lM^{5/2}+O(a^2),
\end{equation}

Note that the duality makes sense if the condition
\begin{equation}
1-\frac{(J^d)^2}{M^2 l^2}>0
\end{equation}
holds, which  leads to an upper bound on NC scale $\Lambda_{NC}$ 
\begin{equation}\label{uvjet}
\Lambda_{NC} < \frac{l}{c \sqrt{M}},\quad c = \frac{64 \pi}{3} \frac{\zeta(2)}{\zeta(3)}.
\end{equation}
 For any macroscopic black hole, eqn. (\ref{uvjet}) easily fulfills the condition that $\Lambda_{NC}\approx\frac{1}{M_{Planck}}$. Moreover, it allows for the limit $a \rightarrow 0$, when we recover the commutative results.

\subsection{Quantum completeness of the  NC BTZ spacetime}

Now, we look at the NC KG equation \eqref{nckg}, where, as it was explained in the previous subsection, we only need to examine the KG equation \eqref{gkg} in the background of a BTZ black hole with spin $J(\Lambda_{NC})$ defined by \eqref{J}. Therefore, the radial part of the NC KG equation, after using the separation of variables $\Phi=R(r)e^{-iEt}e^{im\varphi}$, and change of variables $z=\frac{r^2-r^{2}_{+}}{r^2-r^{2}_{-}}$, is given by
\begin{equation}\label{eom1}
z(1-z)\frac{\d^2 R}{\d z^2}+ (1-z)\frac{\d R}{\d z} + \left(\frac{A}{z}+B\right)R=0,
\end{equation}
where
\begin{equation}\begin{split}
A&=\frac{l^4}{4(r^{2}_{+} - r^{2}_{-})^2}\left(Er_+ -\frac{m}{l}r_{-}\right)^2\\
B&=\frac{-l^4}{4(r^{2}_+ - r^{2}_{-})^2}\left(Er_- -\frac{m}{l}r_+\right)^2
\end{split}\end{equation}

The general solution of \eqref{eom1} can be written as 
\begin{equation}
R(z)=z^{\alpha}F(z)
\end{equation}
where  $F(z)$ satisfies the  hypergeometric equation
\begin{equation}
z(1-z)\frac{\d^2 F}{\d z^2}+\left[c-(1+a+b)z\right]\frac{\d F}{\d z}-abF=0.
\end{equation}
and 
\begin{equation}
a=\alpha+i\sqrt{-B}, \quad b=\alpha-i\sqrt{-B}, \quad c=2\alpha+1
\end{equation}
and 
\begin{equation}
\alpha=i\sqrt{A}.
\end{equation}
In the neighborhood of the outer horizon $r_+$, that is $z=0$, we have two linearly independent solutions
\begin{equation}
R(z)=C_1 z^\alpha F(a,b,c;z)+C_2 z^{-\alpha}F(a+1-c,\ b+1-c,\ 2-c; z)
\end{equation}
Now, the crucial part in our analysis, just like it was illustrated in the previous section, is that we are interested in self-adjoint properties of an operator stemming from \eqref{eom1} when $E\rightarrow \pm i$. In other words, we want to investigate under which condition the equation \eqref{eom1} has NO square integrable solutions for the purely imaginary eigenvalues. Repeating the same procedure as in the previous section, we get that the leading term of interest is
\begin{equation}
\lim_{z\rightarrow 0}\left|R(z)\right|^2 \propto z^{2\Re(\alpha)}
\end{equation}
and this term has to be more divergent than $z^{-1}$ giving us the following condition\footnote{This condition \eqref{con1} was also found in \cite{kumarnormal}, and when this condition holds, the operator in \eqref{gkg} is essentially self-adjoint, rendering the unique time evolution of a scalar field in the BTZ black hole background with intrinsic spin. When the condition is not fulfilled, there is an one-parameter SAE \cite{kumarnormal}.}
\begin{equation}\label{con1}
\Re (\alpha)\leq -\frac{1}{2},
\end{equation}
where $\Re(\alpha)$ stands for the real part of $\alpha$.
When the condition \eqref{con1} is fulfilled, the operator in \eqref{nckg} is essentially self-adjoint and this ensures the unique time evolution for the scalar field in our NC setting. Let us evaluate the condition \eqref{con1} for the physical parameters in the NC setting. Using \eqref{con1}, \eqref{J}, definitions of $r_{\pm}$, and $\alpha=i\sqrt{A(E\rightarrow i)}$, we obtain
\begin{equation}\label{condition1}
\sqrt{M}\leq l\left(1+\Lambda_{NC}\frac{\zeta(2)}{\zeta(3)}\frac{56\pi}{3}\right)
\end{equation}
Comparing with \eqref{condition} we see that the upper bound on the range of the BTZ parameters is enlarged, that is noncommutativity ``smeared out'' the singularity by allowing quantum completeness for a wider range of BTZ parameters.\\

Let as examine now the condition \eqref{uvjet} and take the value $\Lambda_{NC} \propto \frac{l}{c \sqrt{M}}$ as the upper bound for the NC scale. Then by comparing the two conditions \eqref{condition1} and \eqref{uvjet} one can get
\begin{equation}
\sqrt{M}< 1.56 l
\end{equation}
where one witnesses an estimate of  $56$ percent improvement by NC effects. This of course has to be taken with a ``grain of salt'' since the NC scale $\Lambda_{NC}$ can be in principle much lower then the condition \eqref{uvjet} suggests.  \\


\section{Final remarks}
Intuitively, a spacetime singularity is a ``place'' where the curvature ``blows up'' or other ``pathological behavior'' of the metric takes place \cite{Waldbook}. The difficulty in making this notion into a satisfactory one is why some words in the previous sentence are left in quotes. Criterion in which the ``blowing up'' of curvature is defined as an singularity will not include manifold with the so called conical singularity. One can then go toward to the criterion such as geodesic completeness which will detect the ``holes'' of the manifold more successfully, but still there will be a lot of physical situation that will remain geodesically incomplete \cite{Konkowski:2016pgi}.\\

One of these examples is also the spinless BTZ black hole solution \cite{banados}, where there is a conical singularity for $r=0$ that is shielded by the horizon. Unless the spin $J$ in not equal to zero, this spacetime is time-like geodesically incomplete \cite{heno, Cruz}. One can then examine the case of quantum-mechanical completeness in which one is concerned with the uniqueness of the time evolution of test fields. Notice that the quantum completeness depends on the choice of the test field (scalar, fermion, vector, etc.). This is evident from the fact that the effective potential (namely the centrifugal barrier) changes depending on the type of the probe. In \cite{pitelli, Unver} one can see an explicit example in which a singularity can be cured for fermionic probe, but not for the scalar probe. For the case of BTZ spacetime, scalar field is regular at the origin, but we have a problem that it diverges at the horizon. This divergence is due to an infinite redshift and can not be ``transformed away'' by a coordinate transformation since the scalar field is a scalar (same value in all coordinate systems). Therefore one uses the brick wall method \cite{gthooft, thooft1} and examines the dynamics of the field in region $\left\langle r_+, \infty\right\rangle$. Considering just the interval $\left\langle r_+, \infty\right\rangle$ is equivalent to restricting yourself to the outside region of the BTZ black hole, and strictly speaking when considering just this bounded region, this spacetime is geodesically incomplete\footnote{In the same sense as were the various examples from the Introduction.}. However, with the condition \eqref{condition} we see that the spacetime is indeed quantum complete and the evolution of the scalar test field is uniquely defined for all times. We have seen that this situation only gets improved once we introduce NC effects (see equation \eqref{condition1}). One has to stress that considering the BTZ space time just on the bounded region $\left\langle r_+, \infty\right\rangle$ is sufficient to explain/derive Hawking radiation, temperature, entropy, etc. \cite{Kim, Singh}.\\

The study of stability of the nakedly singular negative mass Schwarzschild solution against gravitational perturbations was done in \cite{gibbons}. There the authors used SAE to find a precise criterion for the stability dependance on the boundary conditions. For the BTZ solutions the analogous was done in \cite{pitelli}, where by studying the negative mass parameter they obtained the naked singularity and investigated the quantum completeness. The naked singularity is cured and the space is quantum complete when probed by the test fields. We plan to extend some of these ideas in the NC formalism for the scalar and fermion fields \cite{ncbtz1, ncbtz3}.\\

Semi-classical backreaction was examined in the recent papers \cite{recent, recent1} where it was shown that the backreaction dresses the naked singularity with an event horizon. It would be interesting to see whether these results still hold in the NC setting.\\

In recent papers \cite{Freidel:2017nhg, Freidel:2017wst} the consequences of the noncommutativity of the zero mode sector in toroidal compactification of closed string theory was investigated. From this view point, the zero mode space can be considered as a phase space. It seams that such field theories may not be defined in the UV, but rather in terms of selfdual fixed points. It would be interesting to further investigate this new type of noncommutativity and its implication to the possible new black hole solutions and its singularity structure. \\

\noindent{\bf Acknowledgment}\\
The author is very grateful to Kumar S. Gupta and Andjelo Samsarov for collaboration on the various issues during various stages of this work, and for numerous and fruitful discussions. The author would also like to thank Ivica Smoli\'c for valuable comments and discussions. 
 This work was  supported by the H2020 CSA Twinning project No. 692194, RBI-T-WINNING.\\


\end{document}